\documentclass[aoas,dvips]{arximspdf}

\doi{10.1214/10-AOAS430}
\volume{4}
\issue{4}
\pubyear{2010}
\firstpage{1656}
\lastpage{1656}

\begin{document}
\begin{frontmatter}

\title{Leo and me}
\runtitle{Leo and me}

\begin{aug}
\author[A]{\fnms{Jacob} \snm{Feldman}\ead[label=e1]{feldman@math.berkeley.edu}\corref{}}
\runauthor{J. Feldman}
\affiliation{University of California, Berkeley}
\address[A]{Department of Mathematics\\
University of California, Berkeley\\
Berkeley, California 94720\\
USA\\
\printead{e1}} 
\end{aug}

\received{\smonth{10} \syear{2010}}



\end{frontmatter}

I arrived in Berkeley in 1957, at which time Leo was an Acting
Assistant Professor of Mathematics here. He had recently proven the
``individual ergodic theorem of information theory''---a triumph---and
since this was becoming central to my own interests, it would have been
natural for us to work together. However, Leo's interests shifted to
more applied work, specifically statistics, and he soon moved to UCLA.
So we never became collaborators, but we did became good friends,
especially after 1980 when he returned to Berkeley as a Professor of Statistics.

We had a number of things in common other than mathematics: among them
similar family backgrounds, leftish political views, our connection to Yiddish,
and---for a while---our situation in the world as two divorced men
seeking female companionship.

Leo was quite adventurous, much more so than I. He came up with
proposals which usually I did not go along with. One was to rent a boat
and float along the Sacramento River, drinking, playing cards, and
presumably sweating. Another, which we did do together, was running the
rapids of the American river: Leo, Mary Lou Stagg (whom he later
married), my son Ben and me. That was wonderful, and I have photographs
to prove it. In retrospect, I wish I had gone along with more of his
wild proposals.

We were both born in January 1928, 13 days apart, and for a few years
we would throw large joint parties sometime in between. Sometimes these
would be at his house, sometimes at mine. There were always lots of
people, food, and drink. There would be dancing, and once we even hired
a band. They were fine parties.

Leo's talents, creative energy and imagination spilled out beyond
Mathematics. A few examples: he got interested in elementary education
and spent some time on the Santa Monica school board. He once ran an
ice factory in Mexico. He sculpted, and even had an exhibition of his
work; this was in 1998 at the Nexus Gallery in Berkeley. He took
up---of all things---glass blowing. He designed and supervised the
building of the house in which he and Mary Lou lived during his last
years and in which Mary Lou still lives.

I miss him.


\printaddresses

\end{document}